\documentclass[10pt, letterpaper]{article}
\usepackage{authblk}
\usepackage{hyperref}
\usepackage{amsmath}
\usepackage{amssymb}
\usepackage{textgreek}
\usepackage{graphicx}
\usepackage{epstopdf}
\usepackage{float}
\usepackage{subfigure}
\usepackage[hmargin=2.5cm,vmargin=2.75cm]{geometry}
\usepackage{braket}
\usepackage{wrapfig}
\usepackage{xcolor}
\usepackage{color}
\newcommand{\black}[1]{{\color{black} #1}}
\newcommand{\Rsq}{(\sqrt{3}\times\!\sqrt{3})R30^\circ}

\usepackage[onehalfspacing]{setspace}
\usepackage{bm}
\usepackage[pagewise]{lineno}
\usepackage[
backend=biber,
style=science,
articletitle=true,
maxnames=99,
]{biblatex}

\begin{document}
\frenchspacing

\title{\black{Triggering a} global density wave \black{instability} in graphene via local symmetry-breaking}
\author[1,2]{A. C. Qu}
\author[1,2]{P. Nigge}
\author[3]{S. Link}
\author[1,2]{G. Levy}
\author[1,2,4]{M. Michiardi}
\author[5]{P. L. Spandar}
\author[1,2]{\\T. Matth\'e}
\author[1,2]{M. Schneider}
\author[1,2]{S. Zhdanovich}
\author[3]{U. Starke}
\author[1,2,5]{C. Guti\'errez\thanks{gutierrez@physics.ucla.edu}}
\author[1,2]{\\A. Damascelli\thanks{damascelli@physics.ubc.ca}}
\affil[1]{Department of Physics and Astronomy, University of British Columbia, Vancouver, Canada}
\affil[2]{Quantum Matter Institute, University of British Columbia, Vancouver, Canada}
\affil[3]{Max Planck Institute for Solid State Research, Stuttgart, Germany}
\affil[4]{Max Planck Institute for the Physics of Complex Systems, Dresden, Germany}
\affil[5]{Department of Physics and Astronomy, University of California, Los Angeles, Los Angeles, USA}
\date{}

\maketitle

\textbf{Two-dimensional quantum materials offer a robust platform for investigating \black{the emergence of symmetry-broken ordered phases} owing to the high tuneability of their electronic properties. For instance, the ability to create new electronic band structures in graphene through moir\'e superlattices from stacked and twisted structures has led to the discovery of several \black{correlated} and topological phases. Here we report an alternative method to induce an incipient symmetry-broken phase in graphene at the millimetre scale. We show that an extremely dilute concentration \black{($\bm{<\!0.3\%}$)} of surface adatoms can self-assemble and trigger the collapse of the graphene atomic lattice into a distinct Kekul\'e bond density wave phase, whereby the carbon C-C bond symmetry is broken globally. Using complementary momentum-resolved techniques such as angle-resolved photoemission spectroscopy (ARPES) and low-energy electron diffraction (LEED), we directly probe the presence of this density wave phase and confirm the opening of an energy gap at the Dirac point. We further show that this Kekul\'e density wave phase occurs for various Fermi surface sizes and shapes, suggesting that this lattice instability is driven by strong electron-lattice interactions. Our results demonstrate that dilute concentrations of self-assembled adsorbed atoms offer an attractive alternative route towards designing novel quantum phases in two-dimensional materials.}

\black{A charge density wave (CDW) is a phase of matter that features a spatially modulated electron charge alongside a periodic distortion of the crystal lattice  \cite{gruner_density_2018}.} These spatial modulations break the translational symmetry of the host crystal, \black{resulting in a reconstruction of the Fermi surface that can open energy gaps in the electronic spectrum.} CDW phases have been found in a myriad of systems and can coexist or compete with other correlated phases, such as superconducting and Mott insulating states \cite{monctonNbSe2,tranquada1995evidence,sipos2008mott}. In one dimensional systems, CDW formation is typically thought to be driven primarily by a divergent electronic response to charge modulations with wavevector $\delta q =2k_F$ that connects (or nests) the two points of the Fermi surface \cite{gruner_density_2018,peierls1956quantum}. However, the mechanisms for CDW formation in higher-dimensional systems, where Fermi surface nesting is imperfect, is still highly debated \cite{johannes2008,WeberPhysRevLett.107.107403,zhu2015classification}. 

\begin{figure}[H]
	\centering
	\includegraphics[width=1\linewidth]{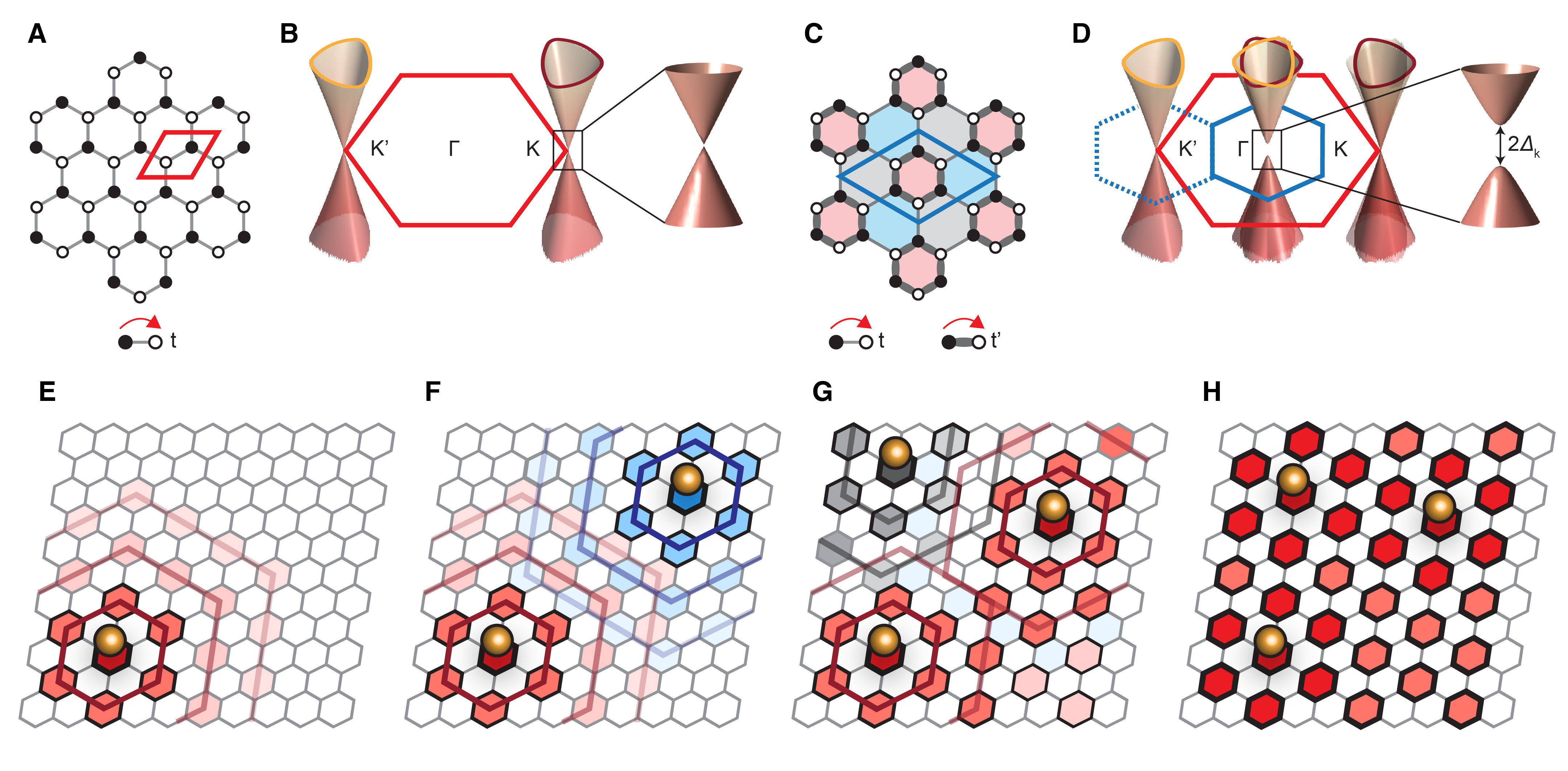}
	\caption{
	\textbf{The Kekul\'e phase and hidden Kekul\'e ordering of lithium on graphene.}
	\textbf{A}, Schematic diagram of pristine graphene. The red diamond indicates the primitive unit cell. All bonds between adjacent atoms are equivalent owing to the same hopping energy, $t$.
	\textbf{B}, Schematic diagram of the graphene band structure composed of two gapless Dirac cones K/K$^\prime$ located at the corners of the Brillouin zone (red hexagon). %Away from the Dirac point, cross sections of the cone begin to warp into triangles.
	\textbf{C}, Schematic diagram of Kekul\'e distorted graphene. The bond symmetry breaks (thin, thick lines) owing to new hopping energies ($t, t^\prime$), forming a $\Rsq$ superstructure and tripling the size of the unit cell (blue diamond). The supercell can be visualized by the red-grey-blue colouring of the three inequivalent hexagonal plaquettes.
	\textbf{D}, The superstructure leads to a smaller Brillouin zone (blue hexagon) and the K/K$^\prime$ Dirac cones being folded to $\Gamma$, as well as a gap opening at the Dirac point. \black{At energies above} the Dirac point, the folded trigonal bands intersect, producing a Star of David pattern. \black{The intensity of folded bands at $K/K^\prime$ is exaggerated for clarity.}
	\textbf{E}, A single adatom on graphene produces Friedel oscillations with a local $\Rsq$ structure on a red site. Bond distortions (thick lines) appear near the adatom.
	\textbf{F}-\textbf{H}, Additional adatoms produce their own Friedel oscillations on out-of-phase sites (blue, gray). The mobile adatoms interact through their Friedel oscillations, preferring to occupy red sites where the oscillations interfere constructively. Hidden Kekul\'e order occurs when a majority of adatoms occupy one of the RGB sites, inducing the Kekul\'e distortion globally.
	\label{fig:intro}
	}
	
\end{figure}

Graphene, the prototypical two-dimensional material, presents an ideal tuneable system for exploring CDW formation: (i) At zero doping, its Fermi ``surface'' consists of two perfectly nested points (K/K$^\prime$) at its Brillouin zone (BZ) corners connected by wavevector $\delta\boldsymbol{q}=\boldsymbol{K}-\boldsymbol{K'}$; (ii) it features strong electron-phonon coupling \black{and soft K-point phonon and plasmon modes} \cite{piscanec_kohn_2004,zhou2008,na2019direct,TudorovskiyPhysRevB.82.073411}; and (iii) its charge density, and thus the size of the Fermi surface and the CDW nesting condition, can be tuned with electrostatic gating or substrate charge-transfer. Owing to these unique properties, and the four-fold degeneracy of its spin and valley degrees of freedom, graphene has been shown to host a plethora of ordered electronic phases. One such phase is the Kekul\'e distortion \cite{chamon_solitons_2000,hou_electron_2007,nomura_field-induced_2009,HerbutPhysRevB.79.085116,weeks_interaction-driven_2010}, a bond density wave phase that has been described as a Peierls-like \cite{peierls1956quantum} lattice instability intrinsic to systems with relativistic dispersions \cite{chamon_solitons_2000,balatsky_peierls_1990,mintmire1992}. Undistorted graphene is composed of a honeycomb lattice of carbon atoms that exhibits six-fold bond symmetry (Fig. \ref{fig:intro}A). \black{Its} low-energy band structure is described by two inequivalent and gapless Dirac cones at K/K$^\prime$ (Fig. \ref{fig:intro}B). In the Kekul\'e phase, the bond symmetry breaks such that the unit cell is tripled in size (thin/thick bonds, Fig. \ref{fig:intro}C). The new commensurate $\Rsq$ bonding pattern can be centred at one of three equivalent carbon hollow sites---distinguished by the use of red-grey-blue (RGB) colour tiling in Fig. \ref{fig:intro}C---and reflects the particular phase of the Kekul\'e order parameter. This supercell causes the previously inequivalent Dirac cones at K/K$^\prime$ to be connected by a reciprocal lattice vector, which leads to a Fermi surface reconstruction that folds each cone to the Kekul\'e BZ center at $\Gamma$ (Fig. \ref{fig:intro}D). \black{Importantly, different topological Kekul\'e phases are determined by the specific broken C-C bond symmetry:  the ``O''-shaped pattern (Fig. \ref{fig:intro}C) opens an energy gap of $2 \Delta_\text{K}$ at the Dirac point \cite{chamon_solitons_2000,hou_electron_2007,gomes_designer_2012} (Fig. \ref{fig:intro}D), while other ``Y''-shaped patterns are gapless \cite{koshinoPhysRevB.90.115207,ren2015,Gutierrez2016, gamayun_valley-momentum_2018,supp}}. When additionally allowing for a possible A-B sublattice symmetry breaking term, $\Delta_\text{AB}$, the total low-energy Kekul\'e dispersion is given by \cite{RyuPhysRevB.80.205319} (Methods):
\begin{equation}
    \varepsilon(k) = \pm\sqrt{\hbar^2 v_F^2 k^2 + \Delta_\text{AB}^2+|\Delta_\text{K}|^2},
\end{equation}
\black{where $\hbar$ is Planck's constant divided by $2\pi$, $v_F$ is the Fermi velocity, and $k$ is the crystal momentum.}

To date there have been few direct experimental observations of the graphene Kekul\'e phase \cite{gomes_designer_2012,Gutierrez2016,liPRBkekule,BaoPhysRevLett.126.206804}. In theory, it can arise through several mechanisms: electron-electron \cite{hou_electron_2007,weeks_interaction-driven_2010,TudorovskiyPhysRevB.82.073411} and electron-phonon interactions \cite{chamon_solitons_2000,nomura_field-induced_2009,kharitonov_phase_2012,classen_instabilities_2014}; from high-density $\Rsq$ adatom superlattices \cite{ren2015,BaoPhysRevLett.126.206804,farjam_energy_2009,sugawara_fabrication_2011,kanetani_ca_2012}; and large biaxial strain \cite{marianetti_failure_2010,lee2011band}. \black{The nature of the Kekul\'e phase is of fundamental interest as it provides an example of spontaneous gap formation via K/K$^\prime$ valley (``chiral'') symmetry breaking \cite{nomura_field-induced_2009,HerbutPhysRevB.79.085116,RyuPhysRevB.80.205319,gusynin2007ac}, and it has been predicted to host topological defects with fractionalized charge \cite{hou_electron_2007}}. Most recently, Kekul\'e lattice instabilities have attracted attention as candidates for the correlated insulating states in twisted bilayer graphene \cite{XuPRB2018Kek,ThomsonPhysRevB.98.075109,daliao2019valence,HuangPhysRevB.101.235140,daliaoPhysRevX.11.011014}. Here we report a controllable method for inducing the Kekul\'e density wave phase in graphene via the deposition of an extremely dilute concentration of lithium adatoms. Using angle-resolved photoemission spectroscopy (ARPES) \cite{damascelli2004probing} we probe the electronic band structure of this density wave phase and directly image the folded Dirac cones at the BZ center, as well as observe the opening of an energy gap of $2\Delta_\text{K}\approx200$ meV at the Dirac point, the two signatures of the Kekul\'e-O phase in graphene. 

We induce the Kekul\'e-O phase \black{(henceforth simply referred to as Kekul\'e)} in graphene through electron-mediated self-assembly of lithium (Li) adatoms. Adatoms on graphene can scatter electrons between valleys K and K$^\prime$, producing unique long-range Friedel oscillations with a hexagonal \black{$\Rsq$} pattern and period determined by scattering wavevectors \black{that satisfy the Kekul\'e nesting condition} $\delta\boldsymbol{q}=\boldsymbol{K}-\boldsymbol{K'}$ \cite{cheianovPRL,Cheianov2009,Cheianov2009_PRB,rutter2007scattering,mallet2012role}. \black{Each Li adatom site thus nucleates a local patch of Kekul\'e density wave order} \black{of radius $\sim$2-5 nm \cite{rutter2007scattering}}, resulting in symmetry breaking between the three previously equivalent RGB Kekul\'e ``colours'' (Fig. \ref{fig:intro}E,F). Other mobile Li adatoms interact with the \black{long-range $\Rsq$ charge modulations}, preferring an arrangement that allows their Friedel oscillations to be in phase (Fig. \ref{fig:intro}G). The constructive interference between several adatom-induced oscillations on a single ``colour'' site produces enhanced charge-density modulations and, through electron-phonon coupling, the emergence of the Kekul\'e distortion throughout graphene---even when the adatoms themselves may be many unit cells apart (Fig. \ref{fig:intro}H) \cite{Cheianov2009,Cheianov2009_PRB}. The adatoms are said to observe hidden Kekul\'e order, since without the RGB colour scheme, the long-range adatom ordering would be obscured \cite{Cheianov2009,Cheianov2009_PRB}.

\begin{figure}[H]
	\centering
	\includegraphics[width=1\linewidth]{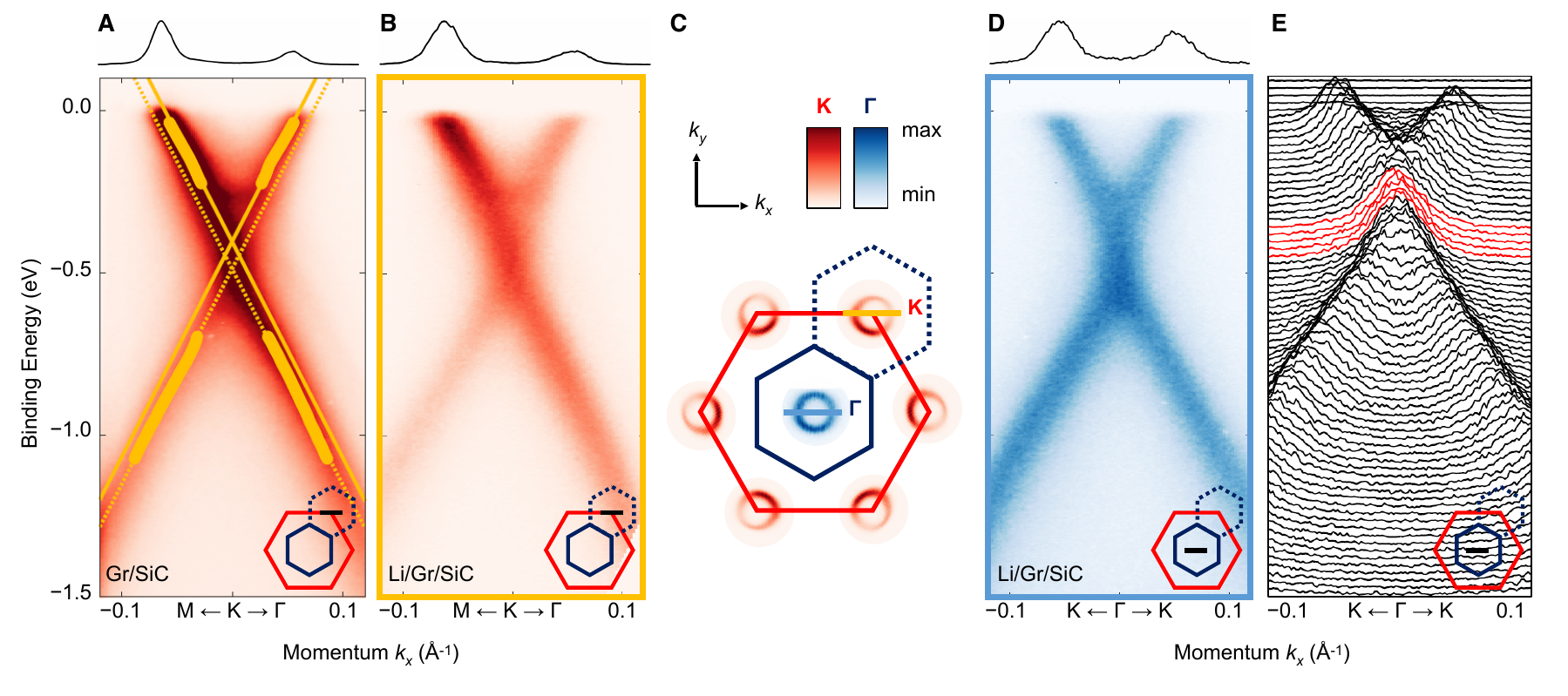}
	\caption{
	    \textbf{Lithium-induced Kekul\'e bond density wave formation in graphene.}
	    \textbf{A}, ARPES spectra of pristine graphene on SiC through the Dirac cone at K (cut indicated by yellow line in \textbf{C}). The Dirac cone displays electron doping due to charge transfer from the substrate, and one branch appears brighter due to matrix element effects. This is shown in the momentum distribution curve (MDC) at the Fermi level (top). MDC peak positions are indicated by thick yellow lines; linear fits through the top (solid lines) and bottom (dotted lines) indicate the presence of an energy gap (see text).
	    \textbf{B}, The same spectra as \textbf{A} after dilute Li deposition. No charge transfer doping from Li is detectable.
	    \textbf{C}, Schematic of the Brillouin zones of pristine graphene (red) and $\Rsq$ Kekul\'e graphene (blue) superimposed on the Fermi surfaces as observed by ARPES (not to scale, see Fig. S9). Locations of the ARPES cuts in \textbf{A}--\textbf{B} and \textbf{D}--\textbf{E} are indicated by the yellow and blue lines, respectively.
	    \textbf{D}, ARPES spectra at $\Gamma$ (cut indicated by blue line in \textbf{C}) after lithium deposition. The two Dirac cones at K/K$^\prime$ are folded to $\Gamma$. Both branches are equally bright, indicating the mixed K/K$^\prime$ character of the bands as shown in the Fermi level MDC (top). \textbf{E}, MDCs of the spectra shown in \textbf{D}. Fits to MDC curves reveal an energy gap at the Dirac point \cite{supp}. The spectra in red lie inside the gap region.
	    \label{fig:arpes}
	}
\end{figure}

In Fig. \ref{fig:arpes}A we show ARPES data from pristine monolayer epitaxial graphene on silicon carbide (Gr/SiC)\cite{Forti_2014} (Methods). \black{ARPES energy-momentum maps} are plotted along the yellow cut in momentum space shown in Fig. \ref{fig:arpes}C, at the graphene K point. Due to charge-transfer from the substrate, the Dirac point is shifted to $\approx$430 meV below the Fermi level $E_F$ \cite{Forti_2014,Zhou2007,Bostwick2007}. The observed intensity asymmetry of the two branches of the Dirac cone in Fig. \ref{fig:arpes}A is due to the experimental geometry \black{and the polarisation of the ultraviolet light source} \cite{gierz2011}. Momentum distribution curves (MDCs) at each binding energy are fit with two Lorentzians, whose peak positions are indicative of the electronic dispersions of the conduction and valence bands (thick yellow lines, \ref{fig:arpes}A). Close to the Fermi energy, the dispersions are well described by linear fits (thin yellow lines, \ref{fig:arpes}A) above and below the Dirac point \black{at $-430$ meV binding energy}. For massless Dirac fermions, the extrapolated fits from the two energy bands would cross at the Dirac point. Such crossing, \black{however}, is not observed. \black{The origin of this apparent energy gap has been attributed to a sublattice symmetry breaking term ($\Delta_\text{AB}$ in Eq. 1) \cite{Zhou2007,nigge2019room} or band renormalization due to electron-plasmon effects \cite{Bostwick2007}. As we describe below, the precise microscopic origin of the pre-existing gap in pristine Gr/SiC is not crucial to our findings, and in either picture our results are qualitatively the same. In order to extract energy gap values, we employ the sublattice symmetry breaking picture which allows us to model our experimental data with a simple tight-binding Hamiltonian (Methods). This analysis yields an energy gap of $2\Delta = (292\pm4)$ meV at the Dirac point \cite{supp}}. Importantly, for pristine graphene near $E_F$, there is no ARPES intensity at $\Gamma$, the BZ centre.

Having confirmed the electronic structure for pristine graphene, we next decorate its surface with Li \black{adatoms}. Due to electron charge transfer from Li, the Dirac point is expected to shift downward in binding energy with increased Li concentration \cite{farjam_energy_2009,sugawara_fabrication_2011,kanetani_ca_2012,Ludbrook2015,supp}. However, we observe only a negligible shift of the Dirac point (below our $20$ meV energy resolution), as shown by a direct comparison of ARPES intensity before (Fig. \ref{fig:arpes}A) and after (Fig. \ref{fig:arpes}B) Li deposition. We conclude that an extremely low Li surface coverage is achieved, with a conservative estimate of \black{less than 0.3\%}, \black{corresponding to an average Li-Li distance of $\geq$2 nm} (Methods). \black{Despite this negligible change in charge density, we observe that the measured energy gap at the Dirac point increases \black{by 20\%} to \black{$2\Delta^\prime = 354\pm2$ meV}.} Since Li \black{adatoms} occupy graphene sublattice symmetric hollow sites (Fig. \ref{fig:intro}E) \cite{farjam_energy_2009}, Li cannot further break A-B sublattice symmetry. \black{Additionally, this gap enhancement after ultra low 0.3\% Li deposition cannot be explained by electron-plasmon effects \cite{supp}.} As we will show below, we attribute this increased gap to the onset of Kekul\'e density wave order. \black{From Eq. (1), Dirac point energy gaps from concomitant sublattice symmetry breaking and Kekul\'e order add in quadrature, \black{$2\Delta^\prime = 2\sqrt{\Delta^2_\text{AB}+\Delta^2_\text{K}}$}, which in our measurements corresponds to $2\Delta_\text{K} \approx 200$ meV.} More striking, and the primary signature of Kekul\'e order, is the appearance of new electronic energy bands centred at the $\Gamma$ point, and directly observed in Fermi surface maps (Fig. \ref{fig:arpes}C). These new bands appear as a replica Dirac cone (Fig. \ref{fig:arpes}D), which we interpret as the Kekul\'e-induced superposition of folded Dirac cones at K and K$^\prime$, as in Fig. \ref{fig:intro}D. This is supported by the momentum-symmetric intensity of the Dirac cone centred at $\Gamma$ (Fig. \ref{fig:arpes}D), which reflects its mixed K/K$^\prime$ character (Fig. \ref{fig:arpes}B) due to contributions from Dirac cones at opposite ends of the graphene BZ (Fig. \ref{fig:arpes}C, red hexagon). By performing a similar MDC analysis on this new Dirac cone at $\Gamma$, we extract a larger energy gap opening of \black{$2\Delta^\prime = 377\pm 2$ meV} in contrast to the gap observed at the BZ corner. A possible explanation for this discrepancy is that measurements at $\Gamma$ exlusively probe areas with the gapped Kekul\'e distorted phase, while those measured at K probe areas with both distorted and undistorted graphene.

\black{Similar band folding and gap opening behaviour has been observed in previous experiments where alkali atoms were intercalated into bilayer graphene at a high density ($33\%$, or 100 times our estimated Li concentration) \cite{BaoPhysRevLett.126.206804,sugawara_fabrication_2011,kanetani_ca_2012}. However, in these experiments, band folding is naturally expected to occur owing to the presence of the uniform $\Rsq$ adatom superlattice potential that explicitly breaks the translational symmetry of graphene. Additionally, this} high adatom coverage strongly electron-dopes graphene, resulting in a significant shift of the Dirac point ($\sim$1 eV) \cite{sugawara_fabrication_2011,Ludbrook2015,linkPRB2019} and the appearance of additional alkali atom-derived electronic bands at $\Gamma$ \cite{farjam_energy_2009,Ludbrook2015}. We detect no such Dirac point shift nor additional Li-derived bands. \black{In contrast, our results are well explained by the global onset of an incipient and phase-coherent Kekul\'e density wave instability driven by the long-range ordering of an extremely dilute ($< 0.3$\%) concentration of mobile Li adatoms} \cite{Cheianov2009,Cheianov2009_PRB}. This is especially evidenced by the particular sharpness of the replica Dirac cone at $\Gamma$ (Figs. 2D--E) over the large spot size of our ultraviolet light source ($\sim$1 mm). This long range phase coherence originates from a majority of mobile Li adatoms occupying a single-coloured Kekul\'e site (Fig. \ref{fig:intro}H); and it is confirmed by additional temperature-dependent measurements (up to 30 K), which found that the folded $\Gamma$-point Dirac bands surprisingly sharpen further as the sample is warmed, indicative of increased hidden Kekul\'e order as more single-coloured Kekul\'e sites are populated by mobile Li adatoms \cite{supp}. \black{Notably, while natural point defects have been shown to nucleate density wave order in bulk materials \cite{melechkoPhysRevLett.83.999,arguello2014visualizing}, this is the first controlled demonstration of dilute extrinsic adatoms inducing a global density wave phase in a two-dimensional material.} 

\begin{figure}[H]
	\centering
	\includegraphics[width=1\linewidth]{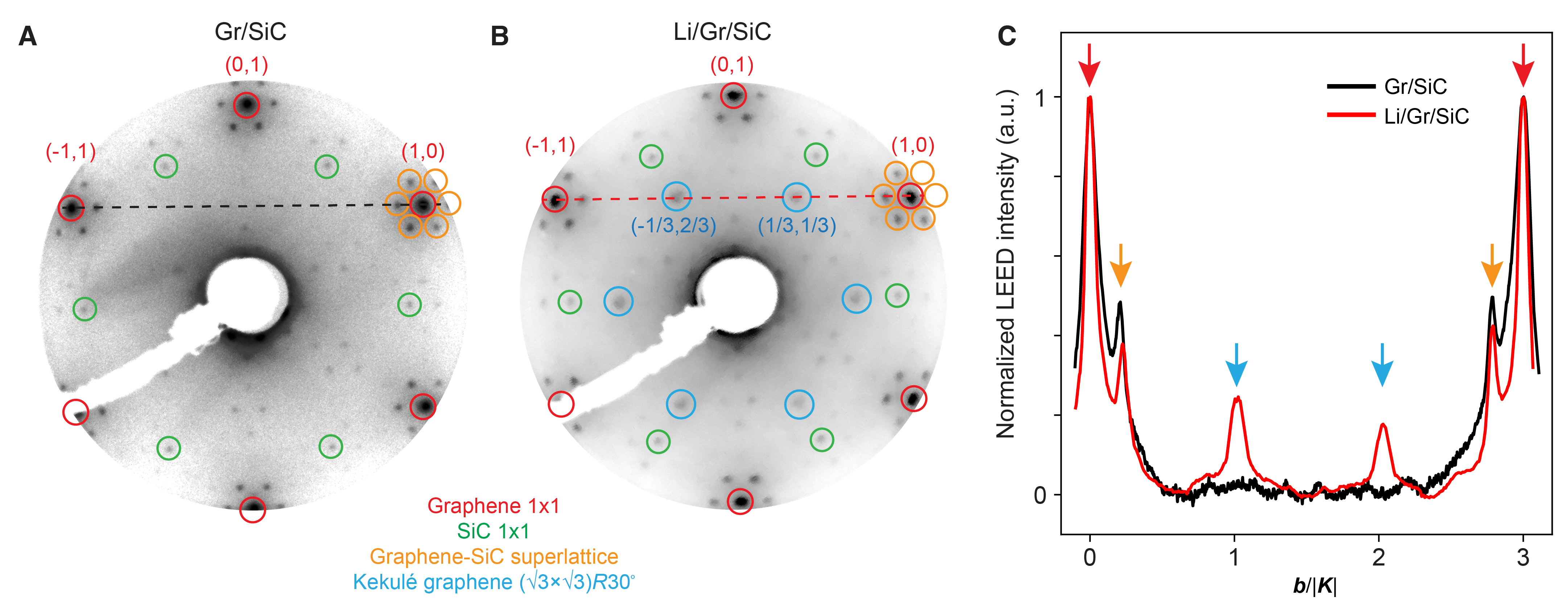}
	\caption{
	\textbf{Low energy electron diffraction (LEED) of graphene in the \black{Kekul\'e phase}.}
	\textbf{A}, LEED pattern measured at 66 eV on pristine Gr/SiC. \black{Peaks are highlighted with circles corresponding to} graphene (red), SiC (green), and the graphene/SiC \black{superlattice} pattern (yellow).
	\textbf{B}, LEED pattern after \black{dilute} lithium deposition at \black{low temperature}. New \black{diffraction} spots corresponding to $\Rsq$ Kekul\'e bond order (blue) are clearly visible.
	\textbf{C}, Line profiles along the black (red) dotted line in \textbf{A} (\textbf{B}) \black{after subtraction of a smooth background \cite{supp}}. The horizontal axis is in units of the K-point wavevector, $|\boldsymbol{K}|$=1.70 \AA$^{-1}$. Diffraction peaks are indicated by arrows using the same colour scheme used in \black{\textbf{B}}. The width of the graphene and graphene/SiC peaks are comparable before and after Li deposition. The new Kekul\'e diffraction peaks (blue arrows) are intense ($\sim$20\%) and as sharp as the graphene Bragg peaks (red arrows), indicating long-range Kekul\'e bond order.}
	\label{fig:leed}
\end{figure}

We note that such direct signatures of density wave order in monolayer graphene are rare \cite{Gutierrez2016,liPRBkekule}. As charge density waves are concomitant with periodic lattice distortions \cite{gruner_density_2018}, a Kekul\'e distortion of the graphene lattice (Fig. \ref{fig:intro}C) should be reflected in diffraction measurements. To further support our claim of a global Kekul\'e distorted phase, we performed \emph{in situ} low temperature low-energy electron diffraction (LEED) measurements (spot size $\sim$1 mm) (Fig. \ref{fig:leed}). Before Li deposition, the pristine surface displays sharp diffraction peaks characteristic of epitaxial graphene grown on SiC (Fig. \ref{fig:leed}A). After dilute Li deposition, new and sharp diffraction peaks appear corresponding to a well-defined and long-ranged $\Rsq$ lattice (Fig. \ref{fig:leed}B). \black{Relative to the atomic graphene Bragg peaks (red arrows, Fig. \ref{fig:leed}C), the new diffraction peaks (blue arrows, Fig. \ref{fig:leed}C) have intensities of $\sim$20\% with similar peak widths \cite{supp}. In light of the ultra-dilute surface coverage of Li adatoms ($<$0.3\%), the intensity and sharpness of these new diffraction peaks points to the presence of a globally well-defined Kekul\'e lattice distortion in graphene.}

\begin{figure}[H]
	\centering
	\includegraphics[width=1\linewidth]{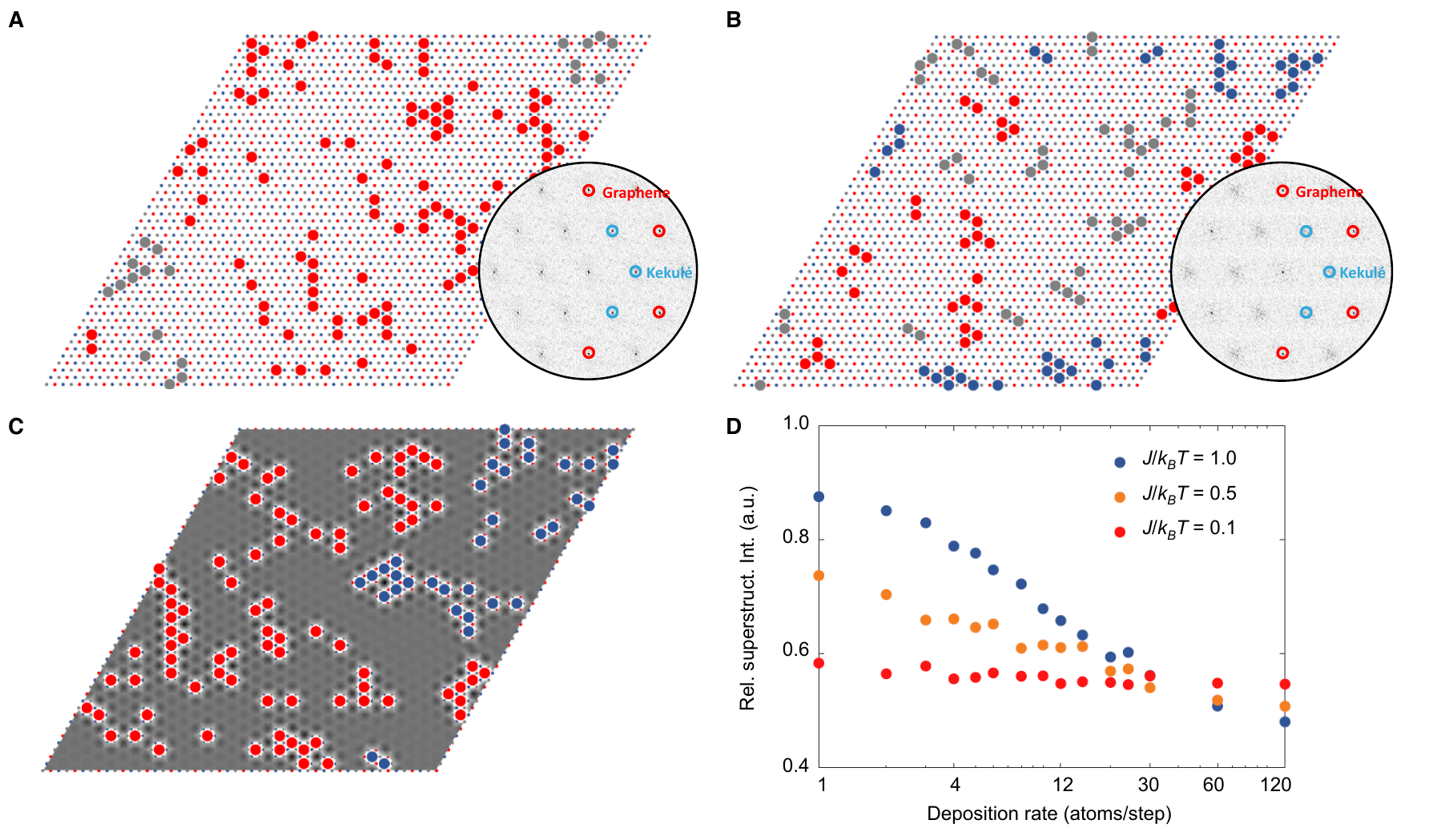}
	\caption{\black{
	\textbf{Kinetic model for hidden Kekul\'e ordering of Li.}
	\textbf{A}, Simulation of 120 adatoms (large dots) added randomly one at a time (slow deposition) to red, grey, or blue (RGB) hollow sites that form three $\Rsq$ mosaic Kekul\'e lattices on graphene. After each Li is added, the adatoms interact through a long range ``ferromagnetic''-like potential (see main text) and the system is evolved thermally \cite{supp}. The formation of a large unicolour (red) Kekul\'e order is observed. Inset: Fourier transform of the simulation, with sharp peaks at the $\Rsq$ Kekul\'e ordering wavevector signifying long-range order. 
	\textbf{B}, Final state of a similar simulation when adatoms are added in a single step (fast deposition) and allowed to evolve thermally. Unicolour patches are much smaller than those from slow deposition in \textbf{A}, and the Kekul\'e peaks in the Fourier transform are broadened, signifying short-range order.
	\textbf{C}, Simulated potential energy landscape originating from the superposition of $\Rsq$ Friedel oscillations from all adatoms. The egg carton-like potential is deepest (dark) near large unicolour regions, where mobile adatoms can become ``trapped'' and further enhance the Kekul\'e order.
	\textbf{D}, Intensity of the Fourier peak corresponding to the $\Rsq$ Kekul\'e lattice (normalized by the intensity at the graphene Bragg peak) as a function of adatom deposition rate, for different values of $J/k_BT$, where $J$ is the adatom-adatom interaction coupling constant (see \cite{supp}), $k_B$ is the Boltzmann constant, and $T$ is temperature. At low temperatures ($J/k_BT\!\geq\!1$), Kekul\'e order is strongest for slow adatom deposition; at high temperatures ($J/k_BT\ll 1$), Kekul\'e order is weak at any deposition rate.
    }}

	\label{fig:toymodel}
\end{figure}

\black{We next present a simple model that captures how the emergence of Kekul\'e order depends on the two key control parameters in our study: sample temperature and Li deposition rate. In our experiments, we find that Kekul\'e order occurs only when (i) Li is deposited with graphene held at low temperature ($<10$ K) and (ii) Li is deposited at a slow rate.} We employ a kinetic hopping model that incorporates the graphene-mediated long-range interaction between Li adatoms on graphene (Fig. \ref{fig:intro}E--H), which is attractive for Li located on Kekul\'e RGB lattices sites of the same ``colour'' and repulsive otherwise \cite{Cheianov2009,Cheianov2009_PRB} (see \cite{supp} for detailed discussion). In the model (Fig. \ref{fig:toymodel}A), Li can hop to neighbouring sites with a Boltzmann probability dependent on temperature and the local Kekul\'e-modulated potential energy landscape created by all other surface Li atoms (Fig. \ref{fig:toymodel}C). We quantify the strength of RGB Kekul\'e order in the system by calculating the magnitude of the $\Rsq$ Fourier component of the Li lattice positions. In agreement with our findings, slower Li deposition (Fig. \ref{fig:toymodel}A) produces coherent long-range Kekul\'e order across the entire sample, while faster deposition (Fig. \ref{fig:toymodel}B) only permits order on the scale of a few sites. \black{By performing several simulations with varying temperatures and deposition rates (Fig. \ref{fig:toymodel}D), we find that long-range Kekul\'e ordering does not form at high sample temperatures. The unique combination of slow deposition at low temperature in our experiments may explain why Kekul\'e order in dilute Li graphene systems has not been reported previously \cite{Ludbrook2015}. However, starting from the ordered phase, higher temperature can initially improve long-range Kekul\'e order via increased Li mobility (Fig. S10) until the kinetic energy eclipses the Kekul\'e confining potential (Fig. \ref{fig:toymodel}C), whence Li hopping becomes random and destroys the ordered phase \cite{supp}.}

\begin{figure}[H]
	\centering
	\includegraphics[width=1\linewidth]{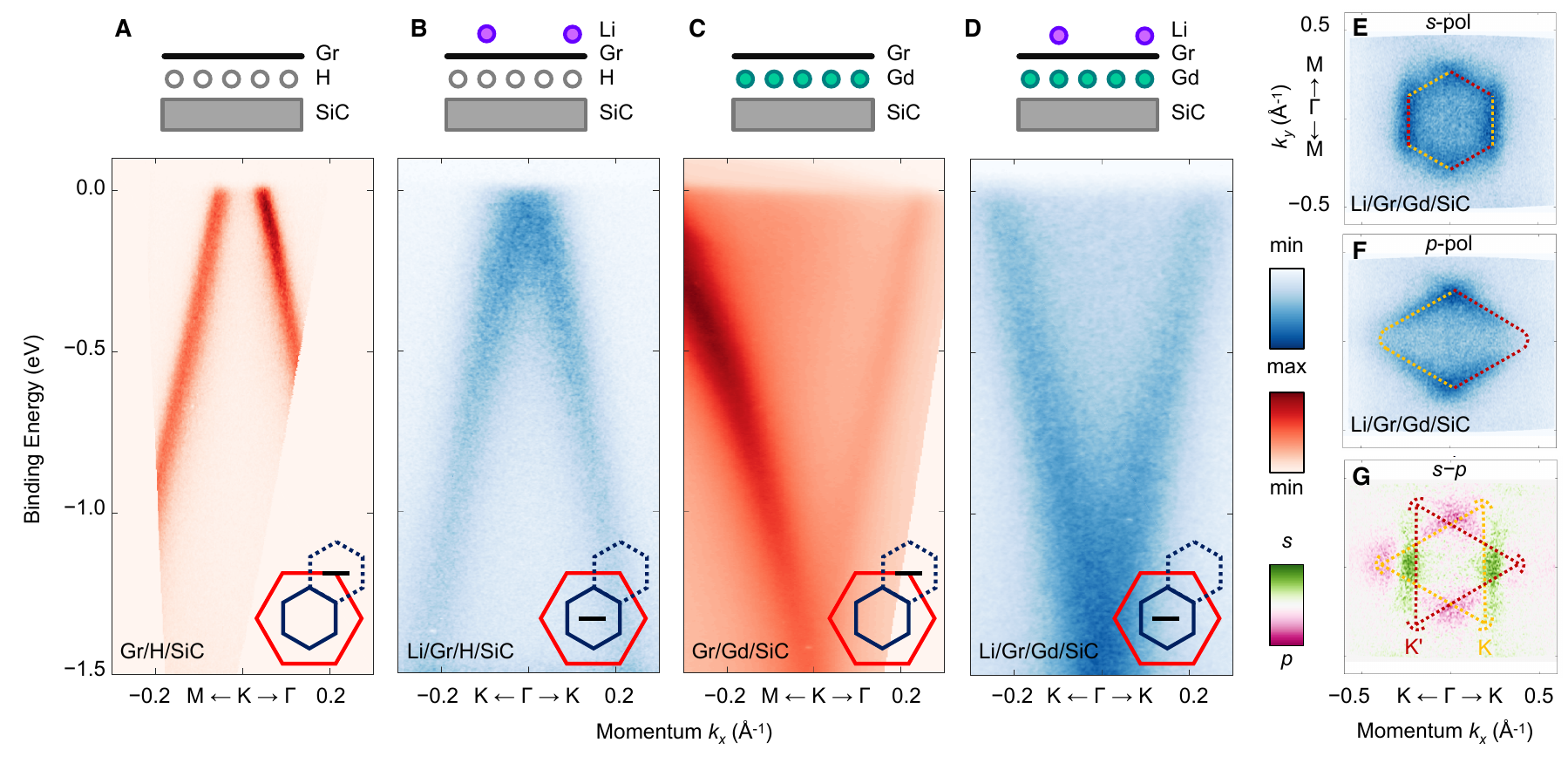}
	\caption{
	    \textbf{Li-induced Kekul\'e phase in different graphene systems.}
	    \textbf{A}, ARPES spectra taken at the K point (cut indicated by inset) on hydrogen-intercalated graphene (Gr/H/SiC) (sample schematic is shown above the cut). The Dirac point is located $\sim$100 meV above $E_F$.
	    \textbf{B}, Spectra at $\Gamma$ on Gr/H/SiC after lithium deposition showing the folded Dirac cone due to Kekul\'e distortion.
	    \textbf{C}, Spectra at K on gadolinium-intercalated  graphene (Gr/Gd/SiC). Graphene is strongly doped with the Dirac point located $\sim$1.6 eV below $E_F$.
	    \textbf{D}, Spectra at $\Gamma$ on Gr/Gd/SiC after lithium deposition showing the folded Dirac cone. The folded Dirac cones in \textbf{B} and \textbf{D} display bands with symmetric intensities, signifying a folding of the two Dirac cones at K/K$^\prime$.
	    \textbf{E}, Fermi surface of Gr/Gd/SiC at $\Gamma$ after Li deposition using $s$-polarized light. A schematic of the bands expected to be visible in this configuration is overlaid in yellow and red dotted lines.
	    \textbf{F}, The same Fermi surface as in \textbf{E} using $p$-polarized light.
	    \textbf{G}, Difference map between \textbf{E} and \textbf{F}. A schematic of the trigonal-warped Fermi surfaces folded from K and K$^\prime$ is overlaid in yellow (K) and red (K$^\prime$) dotted lines.
	    \label{fig:hgd}
	}
\end{figure}

Thus far, our results are well explained by the predictions of hidden Kekul\'e ordering of adatoms \cite{Cheianov2009,Cheianov2009_PRB}. \black{A key parameter for long-range Li order is the graphene doping level:  At finite charge density, additional Friedel oscillations of magnitude $2k_F$ \cite{cheianovPRL,Cheianov2009_PRB,rutter2007scattering,mallet2012role} contribute to the scattered wavevector and disrupt the perfect nesting condition (fig. S9}, $\delta\boldsymbol{q}=\boldsymbol{K}-\boldsymbol{K}' + 2\boldsymbol{k}_F$. \black{When the charge density becomes larger than the density of Li adatoms that nucleate local Kekul\'e order, $n_e\!>\!n_\text{Li}$, the $2k_F$ charge modulations effectively randomize the sign of the interaction between Li adatoms at distances $\ell_\text{Li}\!>1/\sqrt{n_e}$ \cite{Cheianov2009,Cheianov2009_PRB}. Large charge densities are thus expected to destroy any long range Li order, preventing the formation of a global, phase coherent Kekul\'e distortion.} To test this electronic density dependence, we measured two additional epitaxial monolayer graphene samples with different doping levels (Fig. \ref{fig:hgd}). Photoemission spectra from hydrogen (H) and gadolinium (Gd) intercalated graphene-SiC measured at the graphene K point are shown in Figs. \ref{fig:hgd}A and C, respectively. The H-intercalated sample (Gr/H/SiC) is $p$-doped \cite{RiedlPhysRevLett.103.246804}, with the Dirac point shifted to $\sim$100 meV above $E_F$, while the Gd-intercalated sample (Gr/Gd/SiC) is strongly $n$-doped \cite{linkPRB2019} with the Dirac point shifted to $\sim$1.6 eV below $E_F$. Figures \ref{fig:hgd}B and D display ARPES measurements at $\Gamma$ after Li deposition at low temperature and deposition rate. Surprisingly, we again find signatures of Kekul\'e distortion-induced folded K/K$^\prime$ Dirac cones at $\Gamma$. The folded nature of the bands is especially apparent for Gr/Gd/SiC at ultra-high electron density ($n_e >10^{14}$ cm$^{-2}$), where we can directly visualize different portions of the trigonal-warped Fermi surface using different light polarisations \cite{Jung2020} (Fig. \ref{fig:hgd}E--G). Importantly, no \black{ARPES intensity} is present at $\Gamma$ prior to Li deposition.

Despite significant differences in the size and carrier type of the Fermi surfaces, we observe signatures of Li-induced Kekul\'e distortion across all three graphene systems. \black{While understanding the microscopic origin of this Kekul\'e ordering and its robustness to variation in the electron density will require more theoretical work,} the role of Fermi surface nesting in the global lattice distortion can be tentatively ruled out. \black{An alternative, promising mechanism is momentum-dependent electron-phonon coupling, which has been found to drive lattice distortions in other 2D systems \cite{johannes2008,WeberPhysRevLett.107.107403}; indeed, electron-phonon coupling is strong in graphene, and especially so for the K-point $A_{1}^\prime$ ``breathing'' mode phonon \cite{piscanec_kohn_2004,zhou2008,na2019direct}, which has the exact symmetry as the Kekul\'e distortion (Fig. \ref{fig:intro}C) and its dispersion, $\Omega_{A_1^\prime}(\boldsymbol{q})$, displays a strong renormalisation (Kohn anomaly) at the Kekul\'e  wavevector. This hypothesis could be \black{further} explored via time-resolved ARPES experiments, which can extract the mode-projected electron-coupling value \cite{na2019direct} before and after the Kekul\'e distortion is induced.}

\black{Interestingly, Kekul\'e order and other coherent valley-coupled states have been theorized \cite{XuPRB2018Kek,ThomsonPhysRevB.98.075109,daliao2019valence,HuangPhysRevB.101.235140,daliaoPhysRevX.11.011014,angeli2019valley} to explain the correlated insulating and superconducting states in magic angle twisted bilayer graphene \cite{cao_correlated_2018,cao_unconventional_2018,balents2020superconductivity}. However, a microscopic origin of the proposed global intervalley-coupling is yet to be determined. Detailed phonon calculations on relaxed twisted bilayer structures showed that localized Kekul\'e order can emerge near moir\'e AA stacking sites, opening unexpectedly large gaps in the flat band energy spectrum \cite{angeli2019valley}. Defect-induced global Kekul\'e distortion may thus provide a plausible microscopic mechanism for the valley-coupling in twisted bilayer graphene. Most importantly, our results show that even an extremely dilute concentration of atomic-scale defects can initiate the global onset of an incipient density wave instability in pristine monolayer graphene.}

\bigskip

\noindent {\bf Acknowledgements} 

The authors acknowledge useful discussions with J. H. Smet, A. Kogar, R. M. Fernandes, Z.-B. Kang, and I. F. Herbut. {\bf Funding:} This research was undertaken thanks in part to funding from the Max Planck-UBC-UTokyo Centre for Quantum Materials and the Canada First Research Excellence Fund, Quantum Materials and Future Technologies Program. The work at UBC was supported by the Killam, Alfred P. Sloan, and Natural Sciences and Engineering Research Council of Canada's (NSERC's) Steacie Memorial Fellowships (A.D.), the Alexander von Humboldt Fellowship (A.D.), the Canada Research Chairs Program (A.D.), NSERC, Canada Foundation for Innovation (CFI), British Columbia Knowledge Development Fund (BCKDF), and the CIFAR Quantum Materials Program. \black{Work at MPI Stuttgart was supported by the German Research Foundation (DFG) in the framework of the Priority Program No. 1459, Graphene (Sta315/8-2).}  {\bf Author contributions:} A.C.Q. and P.N. performed the ARPES and LEED experiments and analyzed the ARPES data. S.L. and U.S. grew all graphene samples. A.C.Q. performed tight binding calculations. A.C.Q. developed the thermal atomic hopping model and performed calculations with assistance from T. M.  P.L.S. and C.G. analyzed LEED data. A.C.Q., P.N., M.M., M.S., S.Z., and G.L. provided technical support and maintenance for the ARPES setup. C.G. and A.D. supervised the project. A.C.Q. and C.G. wrote the manuscript with input from all authors. A.D. was responsible for overall project direction, planning, and management. {\bf Competing interests:} The authors declare no competing financial interest.

\normalsize

\printbibliography[title=REFERENCES AND NOTES]

%%%%%%%%%%%%%%%%%%%%%%%%%%%%%%%%%%%%%%%%%%%%%%%%%%%%%%%%%%%%%%%%%%%%

\newpage
\section*{Figures}

%%%%%%%%%%%%%%%%%%%%%%%%%%%%%%%%%%%%%%%%%%%%%%%%%%%%%%%%%%%%%%%%%%%%

\end{document}